\PassOptionsToPackage{compress}{natbib}
\documentclass{elsarticle}
\usepackage{graphicx,color}

\usepackage{xcolor}
\definecolor{lcolor}{rgb}{0.5,0,0}
\definecolor{citcolor}{rgb}{0,0.3,0.0}
\usepackage[breaklinks,colorlinks,urlcolor=blue,citecolor=citcolor,linkcolor=lcolor]
{hyperref}

\usepackage{amsmath}

\def\gsim{ \,\, \vcenter{\hbox{$\buildrel{\displaystyle >}\over\sim$}}
 \,\,}

\begin{document}
\begin{frontmatter}
\title{Are the Angular Correlations in $pA$ Collisions due to a
Glasmion or Bose Condensation~?}
\author[baruch,cuny]{A.\ Dumitru}
\author[jyv,hip]{T.\ Lappi}
\author[bnl,rbrc,ccnu]{L.\ McLerran}

\address[baruch]{Department of Natural Sciences, Baruch College, New York, NY 10010, USA}
\address[cuny]{The Graduate School and University Center, The City
  University of New York, 365 Fifth Avenue, New York, NY 10016, USA}
\address[jyv]{Department of Physics, P.O. Box 35, 40014, University of Jyv\"askyl\"a, Finland}
\address[hip]{Helsinki Institute of Physics, P.O. Box 64, 00014 University of Helsinki, Finland}
\address[bnl]{Physics Dept, Bdg. 510A, Brookhaven National Laboratory, Upton, NY-11973, USA}
\address[rbrc]{RIKEN BNL Research Center, Bldg. 510A, Brookhaven National Laboratory,
   Upton, NY 11973, USA}
\address[ccnu]{Physics Department, China Central Normal University, Wuhan, China}

\begin{abstract}
Experiments at the LHC have recently reported results on the angular
asymmetry coefficients $v_n[m]$, for various angular moments $n$ and
orders of cumulants $m$, in high multiplicity $p+Pb$ collisions.  These
coefficients are large, and have both even and odd moments.  We
discuss here some of the implications of these results for our
understanding of the initial state of the collision (Color Glass
Condensate) and for the evolution in the final state (Glasma and
thermalized Quark Gluon Plasma).  We show the Color Glass Condensate
predicts large even moments, $v_n$ with $n$ an even integer.  Odd
moments are generated by final state interactions or fragmentation.
For a multi-particle determination of $v_2[m]$, where m is the number
of particles used to determine the correlation, we argue that if these
coefficients approach equality for large $m$ in high multiplicity
events, this may imply the existence of either solitonic solutions or
Bose condensation either for the JIMWLK action that describes the CGC,
or for the Glasma that might be produced in such a collision.
\end{abstract}
\end{frontmatter}

\section{Introduction}

Recent experiments have shown that there are large angular
correlations in two and four particle correlation functions for
particles far separated in rapidity in heavy ion
collisions~\cite{Ackermann:2000tr,Adler:2002pu,Adler:2003kt,Aamodt:2010pa,ALICE:2011ab,Aamodt:2011by,Voloshin:2012fv,Appelt:2011mw,ATLAS:2012at}, 
$pp$~collisions~\cite{Khachatryan:2010gv},
$pA$~\cite{Chatrchyan:2013nka,Aad:2013fja,Abelev:2012cya} and
dAu~\cite{Adare:2013piz} collisions.  The theory of such flow
coefficients and their applications to hydrodynamic descriptions of
heavy ion collisions is well
developed~\cite{Poskanzer:1998yz,Voloshin:1994mz,Borghini:2001vi}.  
In central heavy ion
collisions, these correlations are usually thought to be
generated by hydrodynamic flow patterns induced by fluctuations in the
transverse position of particles in the heavy ions initiating the
collision~\cite{Andrade:2006yh,Alver:2010gr,Niemi:2012aj}. 
 Although this was
initially thought of as fluctuations of nucleons, subsequent
computations provide a good description based on fluctuations arising
from gluons and quarks, as shown in the IP-Glasma
model~\cite{Gale:2012rq}.  Surely, the existence of such
effects in high multiplicity $pp$ and $pA$ collisions forces one to
have a sub-nucleonic description of the origin of such correlations.
In $pp$~\cite{Dumitru:2008wn,Dusling:2009ni,Dumitru:2010iy,Dusling:2012iga}
and $pA$~\cite{Dusling:2012wy,Dusling:2013oia} collisions, it was
initially thought that final state effects, such as are described by
hydrodynamics, would not be important, and that such correlations if
present would select initial state effects.

Further analysis of the $p+Pb$ data from CERN leads to patterns in the
measured $v_n[m]$ quite similar to those seen in heavy ion collisions,
and suggest that there may be a hydrodynamic interpretation of the
correlation, or at least significant final state
interactions~\cite{Bozek:2012gr,Bozek:2013uha,Werner:2013ipa}.  At this time,
it is not clear whether initial state of final state effect dominate
specific coefficients in various kinematic domains.

In this paper, we discuss effects that might arise either from the
Color Glass Condensate or from early time evolution in the Glasma
phase.  For the CGC, one can provide a viable description of $v_2[2]$,
the second angular moment of the two particle correlation for
both the LHC results on $pp$ and $p+Pb$
scattering\cite{Dumitru:2010iy,Dusling:2012iga,Dusling:2012wy,Dusling:2013oia}. 
We point out
that this prediction also implies significant contribution to all
moments with even values of $n$.  It is however more difficult to
generate odd moments of $v_n[2]$, and we argue that these moments
might arise from final state corrections to the even moments.

Our most interesting observation concerns $v_2[m]$ for various values
of $m$. We find that generically $v_2[m] \sim v_2[2]$.  There is no
suppression due to either factors of coupling or of the number of
colors $N_c$.  There is a numerical coefficient which is of order one,
which also might be negative, but will depend upon $m$, the number of
particles used to compute the correlation.  The experimental data for
high multiplicity $p+Pb$ collisions shows that $v_2[4]$ is quite close
to $v_2[2]$, while they are rather different for central $AA$
collisions.  If it is true that $v_2[m] = v_2[2]$ with small
corrections, this implies that there can be only small fluctuations
from event to event in high multiplicity $p+Pb$ collisions.  This
means that the high multiplicity $p+Pb$ collisions are controlled by
one configuration. This would imply that in either the description of
the Color Glass Condensate or in that of the Glasma, there must be a
single configuration, up to rotational and color zero modes, that
generates the observed angular correlations. This could arise from a
solitonic field configuration, magnetic
vortices~\cite{Dumitru:2013koh}, or a Bose condensation of the gluon
field.  In either the case, for very large systems, a lattice
structure in the transverse plane might form corresponding to
solutions or domains of the condensate.  Such a configuration might or
might not be reflected in instabilities generated early in the
collision~\cite{Mrowczynski:1993qm}.  Because the condensation is
associated with a spin one field, its manifestation would imply the
breaking of rotational invariance, and its evolution would generate
flow-like angular correlations.  This is a very radical result that if
true would imply a major conceptual change in our understanding of
these forms of matter.

\section{The Implications of Measurements of $v_n[2]$}

The LHC experiments have reported results for $v_2[2]$ and $v_3[2]$
for $p+Pb$ collisions.  The results for $v_2[2]$ may be understood in
various hydrodynamic models.  When such hydrodynamic models are
applied to $pp$ and $pA$ collisions, one can question their validity
since very large corrections occur from viscous terms. Also the initial
eccentricity of the matter distribution depends on
uncontrolled assumptions of the distribution of matter in the
transverse plane~\cite{Bzdak:2013zma}, much more so than for larger
collision systems.  Nevertheless it is fair to assume that there will
be some final state interactions at the high multiplicity of particles
where these angular correlations are measured.  After all, in order to
have a saturated Color Glass Condensate, the gluons in the initial
wave function must be strongly interacting, so that for some time
after the collisions, there must also be interactions.  For the moment
we would like however to understand the contributions from initial
state effects of the CGC.

At high transverse momentum, the classical gluon field 
may be expanded in powers of the color sources that generate it,
and the dominant contribution corresponds to one source from each of
the hadrons.  At lower transverse momenta~\cite{Lappi:2009xa,Blaizot:2010kh},
one must include all orders
in the sources from the two hadrons, and the computation may be done
numerically. 
 The
multi-particle correlation functions are computed by taking powers of
the field, $A(q_1) A(-q_1) \cdots A(q_n) A(-q_n)$, and averaging
over the sources which generate the field~\cite{Dumitru:2008wn,Dusling:2009ar,Gelis:2009wh}.

To compute $v_n$  we  average each single $q_i$ with weight 
$e^{\pm i n  \theta}$.  We then extract the coefficients of this expansion. 
For example, for $v_n[2]$, we can denote this as~\cite{Borghini:2001vi}
\begin{equation}
  (v_n[2])^2 = \left< e^{in(\theta_1-\theta_2)} \right>
\end{equation} 
where the angles $\theta_i$ refer to the angles of the two particle
produced in the collision perpendicular to the reaction plane. Higher
order $v_2[m]$ are defined as connected Greens functions so that the
lower order disconnected pieces vanish.  For
example~\cite{Borghini:2001vi},
\begin{equation}
    (v_2[4])^4 = 2\left<e^{2i(\theta_1 -
    \theta_3)}\right> \left<e^{2i(\theta_2-\theta_4)}\right>
  -\left<e^{2i(\theta_1+\theta_2-\theta_3-\theta_4)}\right>~.
\end{equation}
For the quantity $v_2[2]^2$, this Fourier transform generates an
expression in terms of the flow velocities which is of the form of an
operator times its complex conjugate and so is manifestly positive.
On the other hand, for $v_2[4]^4$, the expression is not manifestly
positive.  With the sign chosen above, for a non-zero average flow
field, this expression is positive.  However if the flow field is
small and the expression for $v_2[4]^4$ is fluctuation dominated, any
sign might result.

Implicit in this description are assumptions about the dominant
contributions to these correlation functions.  Note that the quantity
$v_n[1] = \left<e^{in\theta}\right>$, while non zero on an 
event by event basis,
vanishes upon averaging over the orientation of the event plane. 
For a fixed and known impact parameter, one could
define the correlation functions with this mean field value
subtracted out so that connected correlations would isolate
fluctuation contributions.  Then $v_n$ would have been a sum over
background field term plus fluctuations.  Let us denote the angle of
the event plane by $\chi$, and the average over events at fixed
event plane angle to be 
$v_n[1,\chi] = \left<\left< e^{in\theta}\right>\right>(\chi)$. We then have
\begin{equation}
     v_n[1] = {1 \over {2\pi}} \int d\chi ~v_n[1,\chi] = 0~.
\end{equation}
If the mean field term dominated, then
\begin{equation}
    (v_n[m,\chi])^m = {1 \over {2\pi}} \int d\chi~ (v_n[1,\chi])^m  
\end{equation}
and all of the moments coefficients for $v_n[m]$ would be equal for any 
even $m$.

Note however that we would not expect to have a well defined average
angular flow field for small impact parameters.  This means that the
contributions to the various terms for $v_n[m]$ would arise from
fluctuations.  In this case, there is no simple relationship between
the $v_n[m]$, and even the sign of the various terms in the cumulant
expansion might vary.  We will return to this point in the next
section.

To understand the effect of fluctuations compared to an average flow
field, and the effect of rotations of the reaction plane in heavy ion
collisions, it is useful to consider a scalar field theory example.
By analogy, let the Fourier transform $v_2[1,\chi]$ correspond to a
charged scalar field $\phi$.  If there is spontaneous symmetry
breaking of rotational invariance, then $\phi$ will take an
expectation value corresponding to a net flow.  On the other hand, if
one generates such a value of $\phi$ for a finite system, and averages
over many such systems, $\langle\phi\rangle = 0$ by rotational
invariance.  However, quantities such as $\langle\phi
\phi^\dagger\rangle$ which are charge rotationally invariant will be
non-zero, and will give a non-zero $v_2[2]^2$.  Note that for a non
zero value of $\phi$ with our definitions above the quantity
$v_2[4]^4$ is, by analogy, $2\langle\phi \phi^\dagger\rangle \langle\phi
\phi^\dagger\rangle - \langle\phi \phi^\dagger \phi
\phi^\dagger\rangle  = \langle\phi \phi^\dagger\rangle^2$ for $\lambda=0$.
A positive value of the coupling $\lambda$ decreases fluctuations and
thus increases $v_2[4]^4$.  The fluctuations measured with this
cumulant are the non-Gaussian fluctuations associated with a four
scalar interaction.  These are different than the Gaussian
fluctuations that contribute to $\langle \phi \phi^\dagger \rangle$.
While in the region where there are small fluctuations and the
expectation values are dominated by a mean field, there is equality of
$v_2[2]$ and $v_2[4]$, there is no such relation in the fluctuation
dominated region.  In fact, we would generically expect the Gaussian
fluctuations to dominate when fluctuation processes are described by
the central limit theorem. In any case, we conclude that when
fluctuations dominate, it is not enough to think in terms of an
average flow field.

The results of the CGC computation~\cite{Dumitru:2008wn,Dusling:2009ar,Gelis:2009wh}
is that $v_n[2]$ is an even function of $\theta_1-\theta_2$ for the case
of the McLerran-Venugopalan model for the JIMWLK action that generates
the fluctuations in the source terms (see ref.~\cite{Kovchegov:2002nf}
for a computation of $v_2[2]$, for example).  There are however
contributions for all even $n$.  To see that the contribution from
high order moments is large, we need only look at one of the
components which is proportional to
\begin{equation}
C^1_2 = A \left[  \delta^{(2)} (\vec{q}_1 + \vec{q}_2 ) +
  \delta^{(2)} (\vec{q}_1 - \vec{q}_2)  \right] = {1 \over {\pi q_1}}
\delta(q_1-q_2)\sum_n e^{2ni(\theta_1 - \theta_2)} ~.
\end{equation}
In this expression, all of the even moments have an equal weight.  The
other contributions have a more complicated dependence upon the
angular difference $\theta_1 - \theta_2$ but they all in fact have
contributions from all even moments.  Also for higher values of $m$,
the correlation coefficient $v_2[m]$ is not suppressed by factors of
coupling or of $N_c$ and thus there is no reason not to expect
$v_2[m] \sim v_2[2] $.

One might object that the angular contribution above needs to be
smeared out due to particle fragmentation effects, as well as
scattering in the final state.  This is true, but if one smears the
contributions associated with the delta function by a width $\Delta
\theta$ it is easy to see that the moments $v_n[2]$ are not strongly
affected until $n \sim 1/\Delta \theta$.  If one is studying the
correlation for high momentum particles, then we would expect $\delta
\theta \sim Q_s / p$ so this smearing would be small.  For $ p \le
Q_s$, the smearing is of course large, but still one expects
effects of order 1.

How might one generate odd moments?  The most obvious candidate are
particle scattering effects for a finite size media.  Surely if the
inverse saturation momentum is much smaller than the size of the
proton, which is equivalent to $Q_s \gg \Lambda_{\rm QCD}$, then
because particles are strongly interacting with one another in the
saturated CGC, they will also surely interact with one another in the
final state.  This means that if we measure a particle in a two
particle correlation at relatively high momentum it will be biased to
come from a region which is close to the surface of the collision.  If
however, it goes in the backwards direction, it would be suppressed
since it would have the possibility to undergo more interactions that
deflect its momentum.  This effect will result in coefficients of
$v_n[2]$ for all odd moments which may be interpreted as arising from
flow, even though their origin is somewhat different.  The size of the
contribution for odd moments relative to even moments would be
expected to increase as the associated multiplicity and thus the
opacity increases, whereas the effect on the even moments appears even
in the zero opacity limit.

It might also be possible that there are effects not included in the
lowest order expression for the McLerran-Venugopalan model for the
JIMWLK action. These corrections might either come from higher order
terms in the source~\cite{Jeon:2004rk,Jeon:2005cf,Dumitru:2011zz} 
or they might arise from
allowing the sources to exist in a transverse plane which is not
translationally invariant~\cite{Kovner:2010xk}.  The latter case may
be checked in the IP Glasma model; see~\cite{Lappi:2009xa}
for a calculation that does not include fluctuating nucleon coordinates.

In any of the above cases, it is clear that the even moments have
non-zero source contributions  from the initial state.  The odd moments
will have contributions either from final state interactions or higher
order terms which are presumably small for the JIMWLK action.  Seeing
large coefficients for even moments and small for odd might be
interpreted as an indication of the importance of initial state
effects.

In the next section, we will consider a different possible source for
the gluon correlations. We should emphasize at this point that we are
discussing a possible hypothesis to explain the correlation, and that
at this time it is not clear which if any of the scenarios we discuss are
correct.

\section{What Can We Learn from $v_n[m]$?} 

\begin{figure}[h]
\begin{center}
\centerline{
\includegraphics[height=4cm]{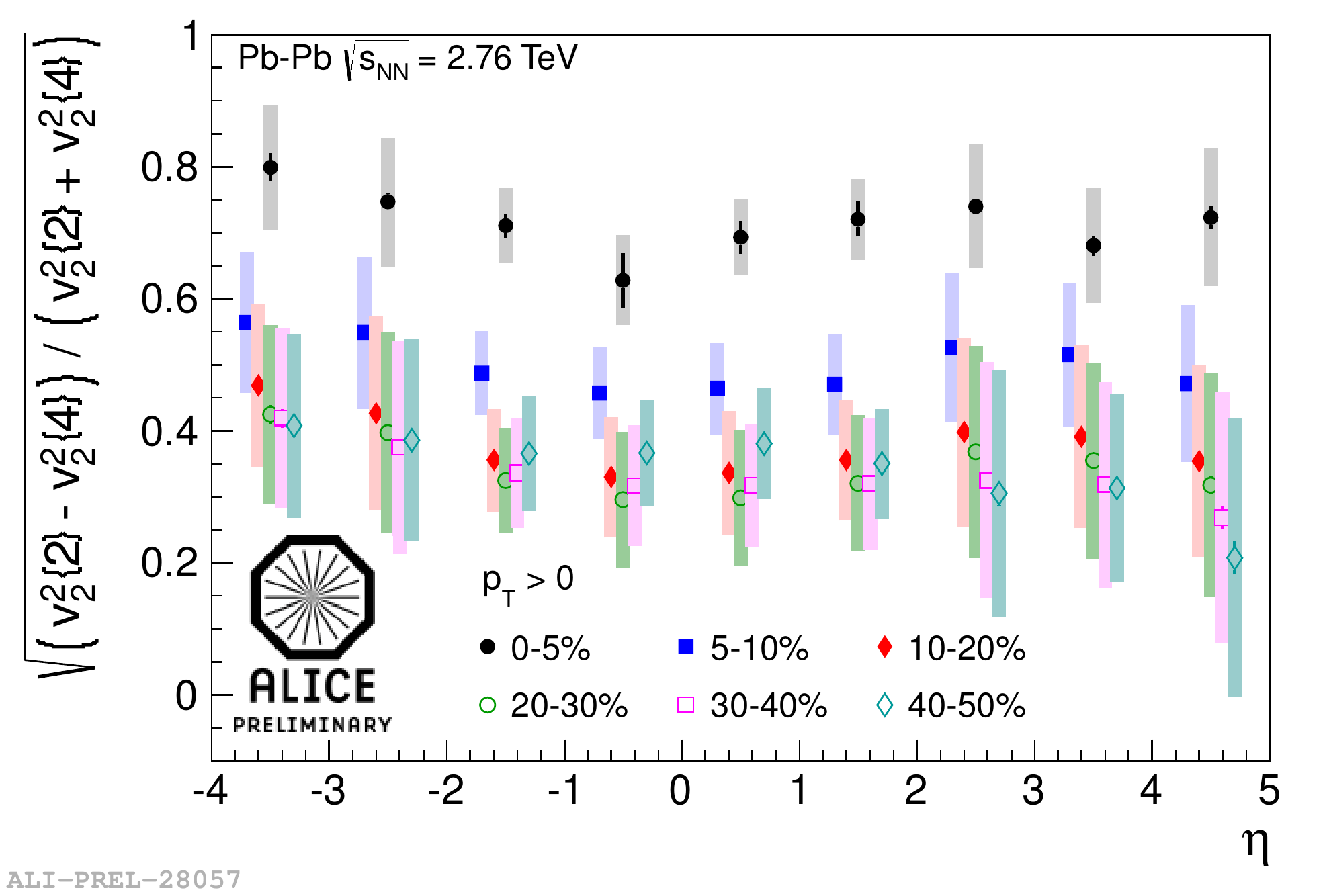}
\includegraphics[height=4cm]{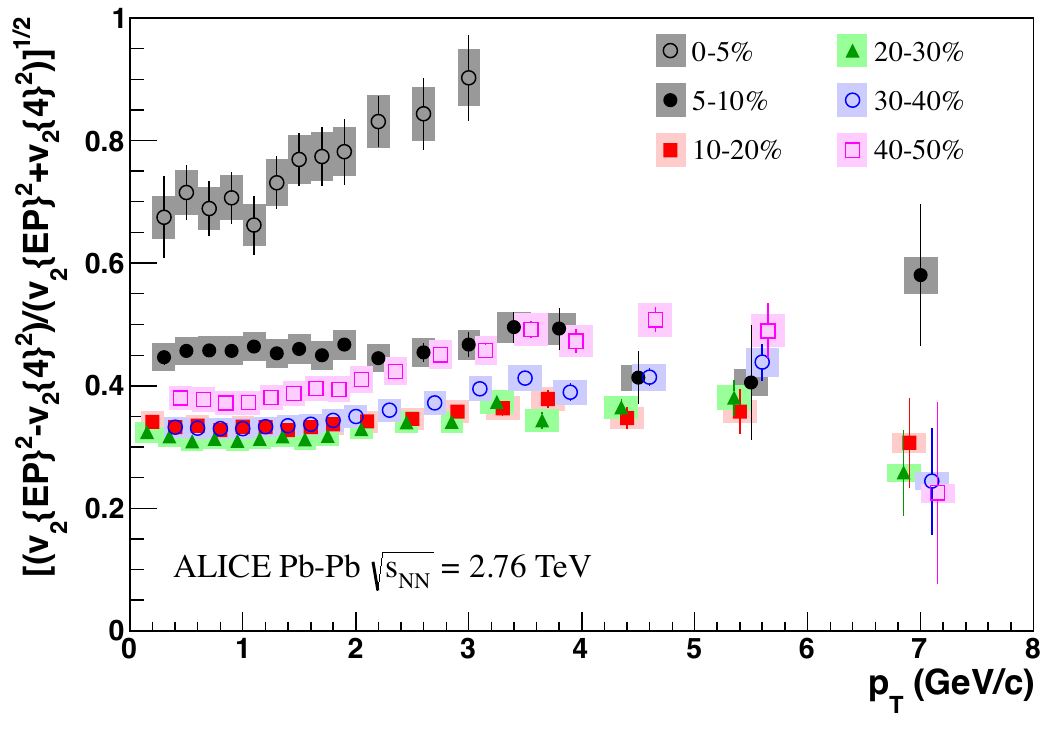}
}
\caption{\label{fluc} Flow fluctuations in the ALICE
  experiment\cite{Voloshin:2012fv}.}
\end{center} 
\end{figure}

\begin{figure}[h]
\begin{center}
\includegraphics[width=8cm]{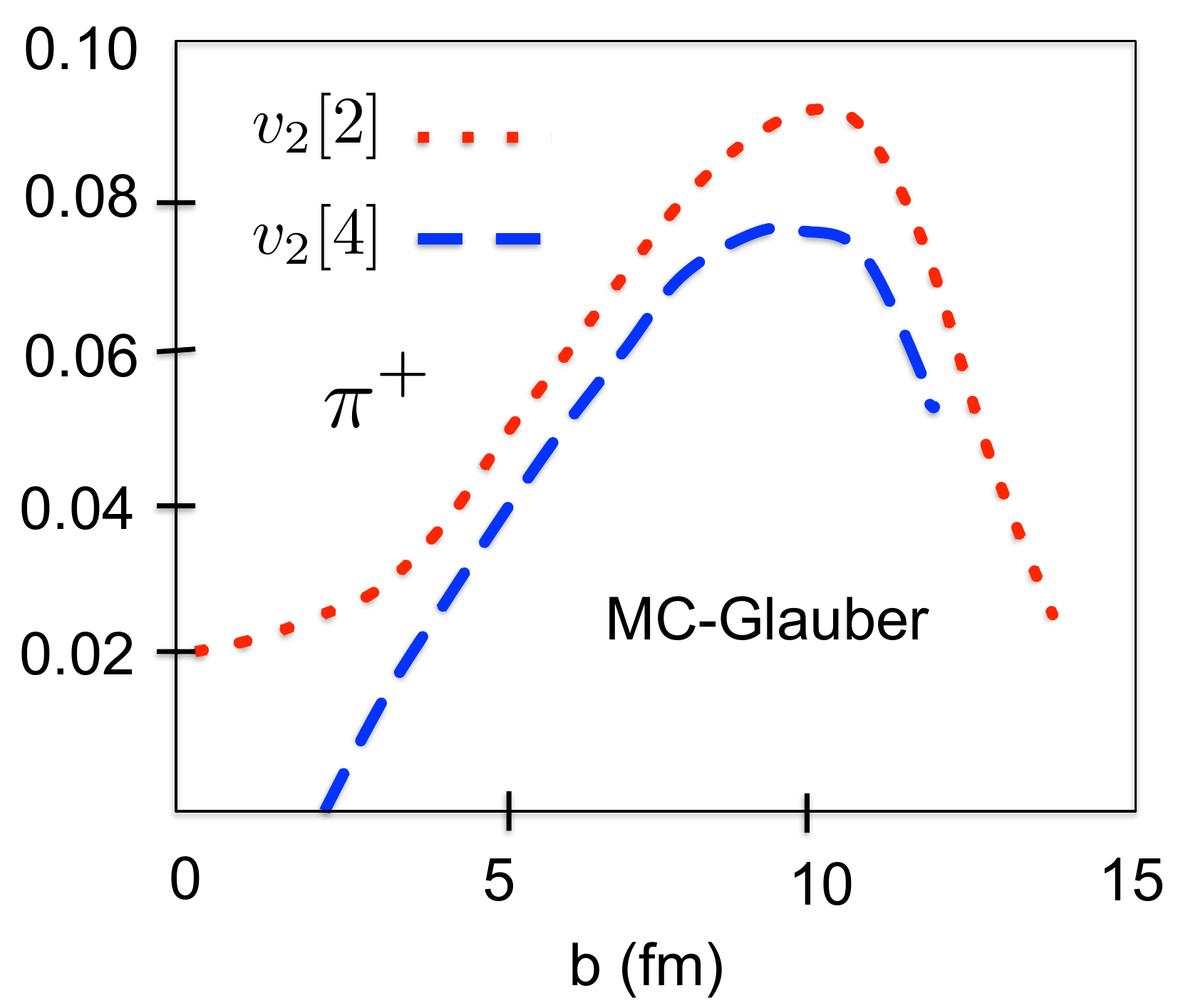}
\caption{\label{heinz} The coefficients $v_2[2]$ and $v_2[4]$ from a
  hydrodynamic computation of nucleus nucleus collisions as a
  function of centrality, based on the result of Ref.~\cite{Qiu:2011iv}.  
}
\end{center} 
\end{figure}

The conventional assumption about the origin of $v_2[m]$ is that it
arises from a mean field term\footnote{That is, from an underlying
  single-particle distribution with a specific, fixed angular emission
  pattern across events.} at a fixed reaction plane angle, that must
be averaged over.  Fluctuations in $v_2[m]$ at a fixed impact
parameter and fixed reaction plane angle are assumed to be small when
one extracts a value of $v_2$ from such a measurement.  However, as we
proceed to smaller and smaller impact parameter, the contribution to
$v_2$ from the mean field should decrease and fluctuations should
become larger.  This can be seen in the data.  In Fig.~\ref{fluc}, the
quantity
 \begin{equation}
 \Delta = \sqrt{{{v_2[2]^2 - v_2[4]^2} \over {v_2[2]^2 + v_2[4]}}}
 \end{equation}
 is plotted as a function of pseudorapidty and of transverse momentum.
This allows us to extract the ratio
\begin{equation}
r = v_2[4]^2/v_2[2]^2 ~.
\end{equation}
For peripheral collisions, we find that $r \sim .8$ and that the flow
coefficients extracted from the higher order contributions are not
strongly affected by fluctuations.  However for central collisions, $r
\sim .2$, and it would appear, as expected, that fluctuation effects
are very important\footnote{This is in agreement with the observation
  by PHOBOS~\cite{Alver:2006wh} that $v_2$ in central heavy-ion
  collisions scales with the so-called ``participant eccentricity''
  which corresponds to the effective elliptic source deformation due to
  fluctuations.}. To see this effect in the context of a
hydrodynamical simulation, consider the computation of $\Delta$ from
ref.~\cite{Qiu:2011iv}, shown in Fig.~\ref{heinz}.  Note that the
four particle cumulant turns negative for very central collisions.

\begin{figure}[h]
\begin{center}\
\includegraphics[width=12cm]{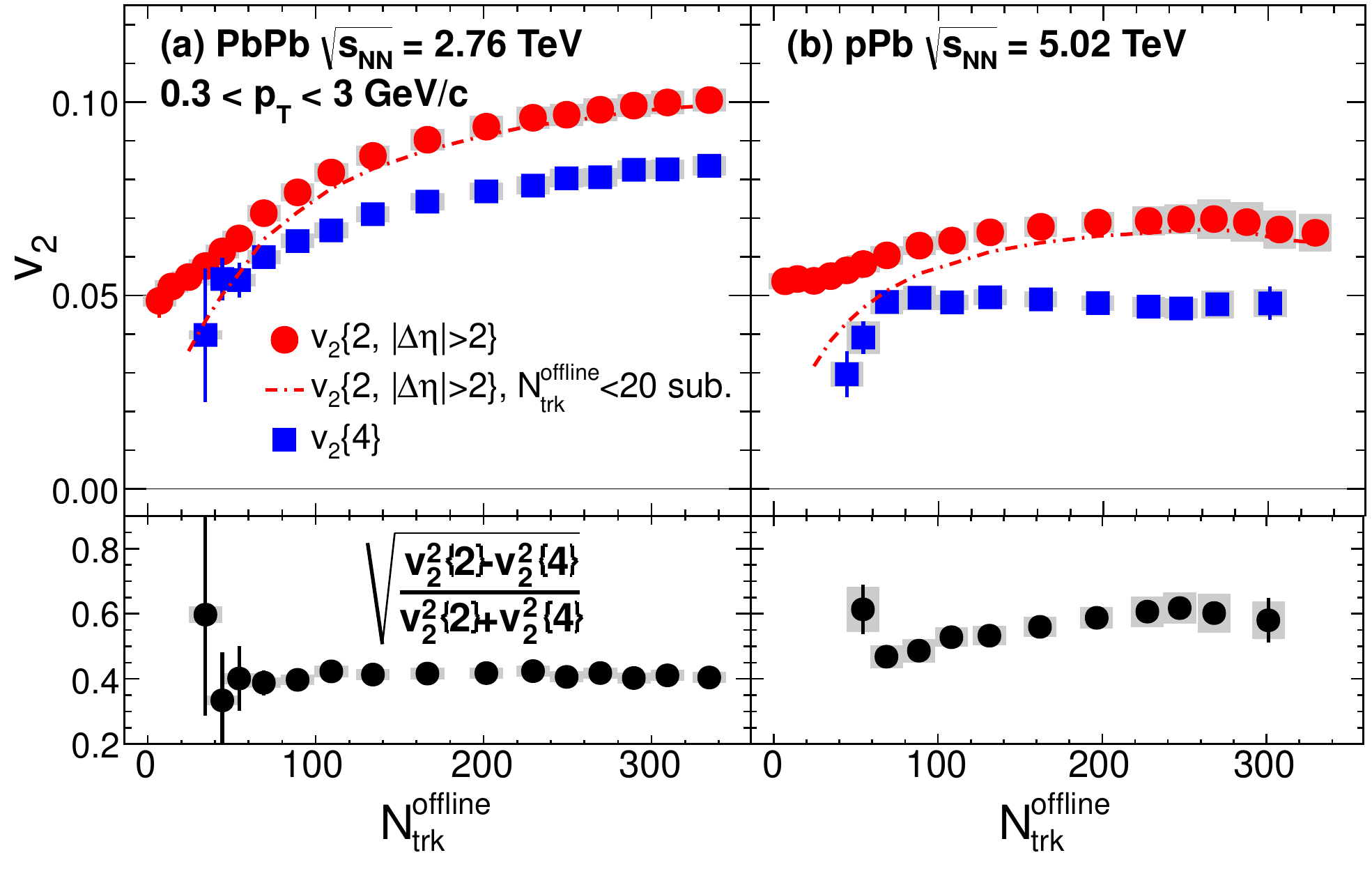}
\caption{\label{ppbfluc} The coefficients $v_2[2]$ and $v_2[4]$ for
  CMS high-multiplicity $p+Pb$ events and for peripheral $Pb+Pb$
  collisions~\cite{Chatrchyan:2013nka}.}
\end{center} 
\end{figure}

The CMS collaboration has extracted the quantity 
$\Delta$ for high multiplicity $p+Pb$
events, shown in Fig.~\ref{ppbfluc}.  They compare their result to
peripheral $Pb+Pb$ collisions.  The remarkable feature of this plot is
that the fluctuations in ``flow'' are small in very high multiplicity
$p+Pb$ collisions.

How can this be?  In peripheral  $Pb+Pb$ collisions one indeed expects
a large mean $v_2$ due to geometry and small relative 
fluctuations around this mean. This is also seen in Monte Carlo Glauber
simulations of the corresponding eccentricities~\cite{Bhalerao:2011yg}.
In contrast, in high multiplicity $p+Pb$ collisions, surely one
is looking at central events.  How can there be a well defined flow if
there is no well defined reaction plane?  In the absence of a clearly
defined reaction plane direction, one could perhaps expect the flow in
$p+Pb$ collisions to be fluctuation dominated. This is, however, not
seen in the data, which indicates a larger mean field contribution.
Nevertheless, the argument for a mean field would be strengthened by a
measurement of $v_2[6]$ and if it was true that $v_2[6] = v_2[4] =
v_2[2]$ to reasonable approximation, then one has to deal with the
issue of a large mean field contribution.

There is a possible way out of the dilemma posed by the 
$p+Pb$ data.  Suppose that instead of
generating the average value for the event by event $v_2[1]$ described
above by impact parameter, that in fact one is generating a classical
solution of either the JIMWLK action or the Glasma equations of
motion.  This could correspond to either a Bose condensate or a
solitonic solution.  More generally the Bose condensate solution might
form a lattice of domains for large systems, and therefore the
distinction between soliton and Bose condensate might become fuzzy.
One does not have an analytic understudying of the fully evolved
JIMWLK equation; it is a two-dimensional theory for the source $\rho$
so it is quite conceivable that there might be a solitonic solution,
or that of a Bose condensate. In the Glasma, it has been argued that
there may be Bose condensation effects due to its highly
coherent nature~\cite{Blaizot:2011xf}.
An explicit calculation of the time development of a solitonic field 
configuration in the Glasma can be found in~\cite{Floerchinger:2013kca}.

Note that since Bose condensation concerns the occupation numbers in the
very infrared modes, finding a gauge invariant definition for 
the condensate in the usual way via occupation 
numbers can be problematic. In numerical simulations one must 
revert to using Coulomb gauge-fixed occupation numbers, or 
alternatively study  explicitly gauge 
invariant observables. The former method was recently employed 
in explicit simulations in the Abelian
Higgs model, where one has seen the emergence of non-trivial solitonic
solution solutions related to vortices and
Q-balls~\cite{Gasenzer:2013era}. An important explicitly gauge invariant 
observable is the spatial Wilso loop.  In fact, it was recently observed that 
the classical field produced
in a collision of two sheets of color charges exhibits area law
scaling of spatial Wilson loops and Z(N) vortex
structure over scales $\gsim 1/Q_s$~\cite{Dumitru:2013koh}.

If there is a soliton or Bose condensation, it solves the problem of
the small amount of fluctuations.  This is because the configuration
corresponds to a stationary phase point of some effective action.  Of
course for a colored vector field, there will be zero modes associated
with spatial rotations and color rotations, but these are treated in
precise analogy to the integration over reaction plane angles
described above for $v_2[1]$, and do not affect the argument about the
smallness of fluctuations when measuring rotationally invariant and
color singlet operators.  Moreover, one expects non-trivial angular
structure because the gluon field is spin-1, and any expectation value
for the gluon field will break both color rotational and spatial
rotational invariance. Even if a potential solitonic solution decays fast,
as has been argued in~\cite{Kurkela:2012hp}, the decay products could 
still retain some sensitivity to the coherent nature of the initial
stage, and the consequences for experimental observables could be similar.

One can imagine two rather different scenarios depending on how such
solitonic or condensate solutions might arise.  If it is from the
JIMWLK action and it is an initial state property then such an ordered
solution might extend over the entire transverse plane.  We would
expect structure on a scale of the order of the saturation momentum,
which was replicated up to the scale of the size of the nucleus.  In
this case, the transverse system should look like some sort of two
dimensional lattice.  If, on the other hand, the structure arose in
the Glasma, its transverse extent would be limited by causality, and
it might only appear in small systems such as $pp$ or $pA$ collisions.
While in peripheral $AA$ collisions the relative flow fluctuations are
small because the flow itself is large, this might explain why 
they can also be small in central $pA$ collisions.

 \section{Summary}
 
We have proposed a very radical solution to the origin of the large
angular ``flow'' correlations observed for central $p+Pb$
collisions. Azimuthal correlations could arise from a large mean
field rather than being fluctuation dominated. However, rather than
assuming a well developed hydrodynamic flow field as the underlying
mechanism for the fixed (across events) single-particle distribution
we point out the possible existence of a dominant classical stationary
phase point of the effective action. Clearly, a measurement of
$v_2[6]$ would be a strong confirmation of our hypothesis if it turned
out that $v_2[6] = v_2[4]$ with good accuracy.
 
In the case of condensation, we expect some domain structure on a size
scale of order the saturation momentum~\cite{Dumitru:2013koh}.  It
would be difficult to imagine color structures forming which have long
range.  Such domain like structures might be thought of as solitons,
and the resulting lattice a solitonic lattice.  For a large system
like a nucleus, this lattice would be no doubt disordered on some size
scale associated with causality.  Such a hypothetical
localized solution could be either a domain in the condensate, or perhaps even
a true solitonic solution that we call the Glasmion.  Clearly, such structures
must be sought in numerical simulations of Glasma
evolution~\cite{Berges:2013eia,Epelbaum:2013waa,Gelis:2013rba}.

\section*{Acknowledgements}
The research of LM is supported under DOE Contract
No.\ DE-AC02-98CH10886.  LM thanks J.\ Berges, J.\ Pawlowski, B.\ Schenke
and R.\ Venugopalan for informative discussions issues related to the
results of this paper.  He especially thanks Juergen Schukraft and
Raju Venugopalan for provocative statements about higher order flow
coefficients and their possible implications.  He was supported as a
Hans Jensen Professor of Theoretical Physics in the Theoretical
Physics Institute during the time this research was initiated and
completed. 
AD thanks T.\ Kodama for sharing his insight on coarse graining in
hydrodynamics; he gratefully
acknowledges support by the DOE Office of Nuclear
Physics through Grant No.\ DE-FG02-09ER41620 and from The City
University of New York through the PSC-CUNY Research Award Program,
grant 66514-0044.
TL is supported by the Academy of Finland, projects
133005, 267321 and 273464.

\bibliography{spires}
\bibliographystyle{h-physrev4mod2}

\end{document}